\documentclass[useAMS,natbib]{mn2e}
\usepackage{epsf,graphics}
\usepackage{rotating,times,pictex,graphicx,latexsym,color,longtable}
\def\deg{$^\circ$}        
\def\arcsec{$"$}        
\def\arcmin{$'$}        

\newcommand{\mic}{\,$\mu$m}

\newcommand{\htwo}{H$_2$}

\newcommand{\brg}{Br$\gamma$}

\title[The UKIRT Widefield Infrared Survey for H$_2$]
{UWISH2 -- The UKIRT Widefield Infrared Survey for H$_2$}

\author[D.~Froebrich et
al.]{\parbox{\textwidth}{D.~Froebrich$^{1}$\thanks{E-mail: df@star.kent.ac.uk},
C.J.~Davis$^{2}$\thanks{E-mail: c.davis@jach.hawaii.edu}, G.~Ioannidis$^{1}$,
T.M.~Gledhill$^{3}$, M.~Takami$^{4}$, A.~Chrysostomou$^{2}$, J.~Drew$^{3,5}$,
J.~Eisl\"offel$^{6}$, A.~Gosling$^{7}$, R.~Gredel$^{8}$, J.~Hatchell$^{9}$,
K.W.~Hodapp$^{10}$, M.S.N.~Kumar$^{11}$, P.W.~Lucas$^{3}$, H.~Matthews$^{12}$,
M.G.~Rawlings$^{13,14}$, M.D.~Smith$^{1}$, B.~Stecklum$^{6}$, 
W.P.~Varricatt$^{2}$, H.T.~Lee$^{4}$, P.S.~Teixeira$^{15}$, C.~Aspin$^{10}$, 
T.~Khanzadyan$^{16}$, J.~Karr$^{4}$, H.-J.~Kim$^{17}$, B.-C.~Koo$^{17}$,
J.J.~Lee$^{18}$,  Y.-H.~Lee$^{17}$, T.Y.~Magakian$^{19}$,
T.A.~Movsessian$^{19}$,  E.H.~Nikogossian$^{19}$, T.S.~Pyo$^{20}$,
T.~Stanke$^{15}$ \vspace{0.4cm}} \\ $^{1}$Centre for Astrophysics \& Planetary
Science, The University of Kent, Canterbury, Kent CT2 7NH, UK \\ $^{2}$Joint
Astronomy Centre, 660 North A'ohoku Place, University Park, Hilo, HI 96720, USA
\\ $^{3}$Centre for Astrophysics Research, University of Hertfordshire, College
Lane, Hatfield AL10 9AB, UK \\ $^{4}$Institute of Astronomy and Astrophysics,
Academia Sinica, PO Box 23-141, Taipei 10617, Taiwan \\ $^{5}$Physics
Department, Imperial College London, Exhibition Road, London SW7 2AZ, UK \\
$^{6}$Th\"uringer Landessternwarte, Sternwarte 5, 07778 Tautenburg, Germany \\
$^{7}$Department of Astrophysics, University of Oxford, Keble Road, Oxford OX1
3RH, UK \\ $^{8}$Max Planck Institut f\"ur Astronomie, K\"onigstuhl 17, 69117
Heidelberg, Germany \\ $^{9}$School of Physics, University of Exeter, Stocker
Road, Exeter EX4 4QL, UK \\ $^{10}$Institute for Astronomy, University of
Hawai'i at Manoa, 640 N. Aohoku Place, Hilo, HI 96720, USA \\ $^{11}$Centro de
Astrofísica da Universidade do Porto, Rua das Estrelas, 4150-762 Porto, Portugal
\\ $^{12}$National Research Council Canada, Herzberg Institute of Astrophysics,
5071 West Saanich Rd, Victoria, BC, V9E 2E7, Canada \\ $^{13}$Joint ALMA
Observatory, Alonso de C\'ordova\,3107, Vitacura 763-0355, Chile \\ 
$^{14}$European Southern Observatory, Santiago Office, Alonso de
C\'ordova\,3107, Vitacura, Casilla 19001, Chile \\ $^{15}$European Southern
Observatory, Karl-Schwarzschild-Stra{\ss}e 2, D-85748 Garching bei M\"unchen,
Germany \\ $^{16}$Max Planck Institut f\"ur Radioastronomie, Auf dem H\"ugel 69,
53121 Bonn, Germany \\  $^{17}$Department of Physics and Astronomy, Seoul
National University, Seoul 151-747, Korea \\ $^{18}$ Korea Astronomy and  Space
Science Institute, Daejeon 305-348, Korea \\ $^{19}$Byurakan Astrophysical
Observatory, 378433 Aragatsotn reg., Armenia \\ $^{20}$Subaru Telescope,
National Astronomical Observatory of Japan, 650 North A'ohoku Place, Hilo, HI
96720, USA}   \date{Accepted ..... Received ..... ; in original form .....}
\pagerange{\pageref{firstpage}--xxx}\pubyear{2010}

\begin{document}
\maketitle
\label{firstpage}

\begin{abstract} We present the goals and preliminary results of an unbiased,
near-infrared, narrow-band imaging survey of the First Galactic Quadrant
(10\deg\,$< l <$\,65\deg\ ; $-1.3$\deg\,$< b <$\,+1.3\deg). This area includes
most of the Giant Molecular Clouds and massive star forming regions in the
northern hemisphere. The survey is centred on the 1-0\,S(1) ro-vibrational line
of \htwo, a proven tracer of hot, dense molecular gas in star-forming regions,
around evolved stars, and in supernova remnants. The observations complement
existing and upcoming photometric surveys (Spitzer-GLIMPSE, UKIDSS-GPS,
JCMT-JPS, AKARI, Herschel Hi-GAL, etc.), though we probe a dynamically active
component of star formation not covered by these broad-band surveys. Our
narrow-band survey is currently more than 60\,\% complete. The median seeing in
our images is 0.73\arcsec. The images have a 5\,$\sigma$ detection limit of
point sources of K$\sim$18\,mag and the surface brightness limit is
10$^{-19}$\,W\,m$^{-2}$\,arcsec$^{-2}$ when averaged over our typical seeing.
Jets and outflows from both low and high mass Young Stellar Objects are
revealed, as are new Planetary Nebulae and - via a comparison with earlier
K-band observations acquired as part of the UKIDSS GPS - numerous variable
stars. With their superior spatial resolution, the UWISH2 data also  have the
potential to reveal the true nature of many of the Extended Green Objects found
in the GLIMPSE survey. \end{abstract}


 %
 %
 %
 %
 %

\section{Introduction}
 
Feedback from star formation, particularly massive star formation, has a radical
impact on the nature of the interstellar medium (ISM) in galaxies. Outflows from
protostars and radiated energy from high-mass young stars heat, excite, modify
the chemistry of and may provide the turbulent motions in Giant Molecular Clouds
(GMCs). Ultimately, massive stars also enhance metal abundances in the ISM.
Understanding the formation of stars, particularly massive stars, is thus of
crucial importance.

To help us better understand the dynamical processes associated with massive
star formation, but also to search for other line-emitters (Supernova remnants,
Planetary nebulae, etc.) along the Milky Way, we are conducting an unbiased
survey of the Spitzer Space Telescope GLIMPSE-North portion of the Galactic
Plane in \htwo\ 1-0\,S(1) emission at 2.122\,$\mu$m. \htwo\ observations
highlight regions of shocked or fluorescently excited molecular gas (T $\approx$
2000\,K, n$_{\rm H_2} >$\,10$^3$\,cm$^{-3}$) and thus trace outflows and jets
from embedded young stars, but also the radiatively excited boundary regions
between massive stars and the ISM.

Our unique survey -- dubbed the UKIRT Wide Field Infrared Survey for \htwo\, or
UWISH2 -- complements the existing UKIRT Infrared Deep Sky Survey (UKIDSS) of
the Galactic Plane (the UKIDSS GPS; Lawrence et al. \cite{2007MNRAS.379.1599L},
Lucas et al. \cite{2008MNRAS.391..136L}), as well as the Isaac Newton Telescope
Photometric H$\alpha$ Survey (IPHAS: Drew et al. \cite{2005MNRAS.362..753D}),
the Herschel Infrared Galactic Plane Survey (Hi-GAL: Molinari et al.
\cite{2010PASP..122..314M}), the planned James Clerk Maxwell Telescope
Submillimeter Common User Bolometer Array-2 (SCUBA-2) Galactic Plane Survey (the
JPS), the Very Large Array 5\,GHz "CORNISH" survey (Purcell et al.
\cite{2008ASPC..387..389P}), the AKARI mid-infrared all sky survey (Ishihara et
al. \cite{2010A&A...514A...1I}), and of course the mid-infrared GLIMPSE survey
with Spitzer (Benjamin et al. \cite{2003PASP..115..953B}, Churchwell et al.
\cite{2006ApJ...649..759C}) as well as the MIPSGAL survey (Carey et al.
\cite{2009PASP..121...76C}). Other surveys which are covering a large fraction
of our field are the Bolocam Northern Galactic Plane Survey at 1.1\,mm (BGPS,
Rosolowsky et al. \cite{2010ApJS..188..123R}, Aguirre et al.
\cite{2010arXiv1011.0691A}), the APEX Telescope Large Area Survey of the Galaxy
at 870\,$\mu$m (ATLASGAL, Schuller et al. \cite{2009A&A...504..415S}), the
Millimetre Astronomy Legacy Team 90 GHz Survey (MALT\,90), the Galactic Ring
Survey (GRS, Jackson et al. \cite{2006ApJS..163..145J}) and the Methanol
Multibeam Survey (MMB, Green et al. \cite{2007IAUS..242..218G}). 

Together the GLIMPSE, UKIDSS-GPS, JPS, Hi-GAL, CORNISH and other surveys provide
a near-complete census of star formation, detecting cool, pre-stellar cores, hot
cores, low and high-mass protostars (Class\,0/I), pre-main sequence objects
(Class\,II/III -- T\,Tauri and Herbig Ae/Be stars) and H{\small II} regions
associated with intermediate and high-mass young stars. The longer wavelength
surveys will also map the associated dust distribution. Outflows from low and
high-mass protostars, as well as H{\small II} regions, Photo-Dissociation
Regions (PDRs), post-Asymptotic Giant Branch (post-AGB) stars and planetary
nebulae (PN) are detectable by the UWISH2 survey and, in many cases, are
resolved out to distances of at least 5\,kpc.

\begin{figure*}
\includegraphics[width=6cm]{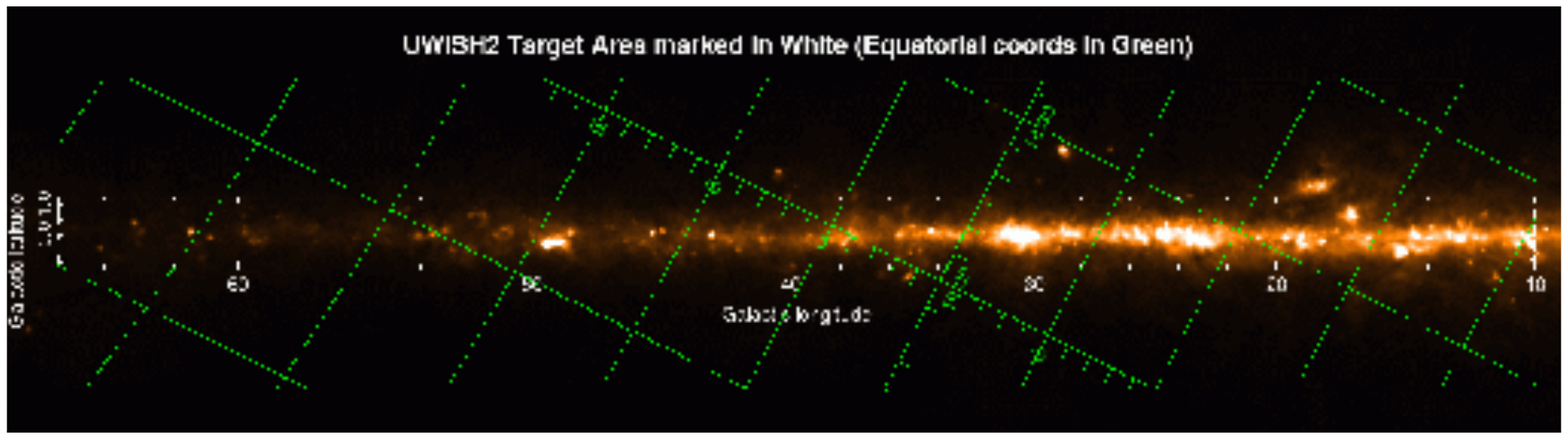} \\
\includegraphics[width=6cm]{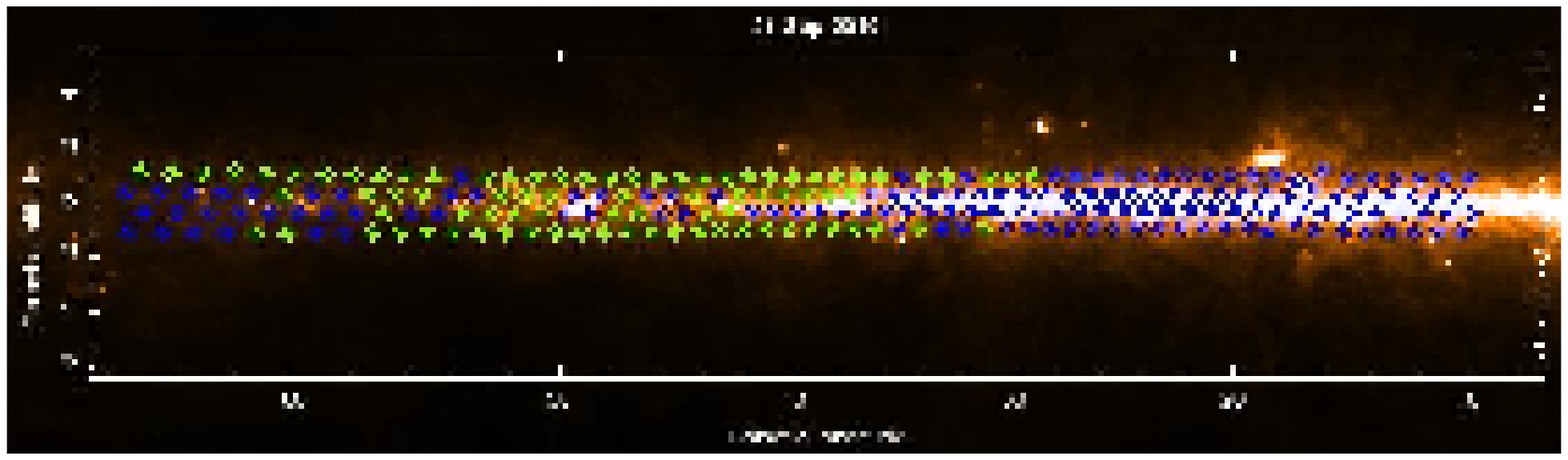}
\caption[]{\label{area} {\bf Left:} An overview of the UWISH2 target region in
the Galactic Plane. The image shows the IRAS dust emission at 100\mic\ with the
Equatorial coordinate system overlayed in green and the target region in white.
{\bf Right:} A map showing the completeness of the UWISH2 survey observations as
of 21 September 2010 (i.e. after our observing run in Semester 2010B). Blue
symbols mark observed tiles; green symbols mark remaining tiles. As of September
2010 the survey is 62.9\,\% completed. The symbols mark the four positions
observed within each tile. Note that there are no gaps between adjacent tiles.}
\end{figure*}

\section{Target Area}

We have selected a region within the Galactic Plane, known as the First Galactic
Quadrant, that has already largely been mapped by WFCAM through broad-band J, H,
and K filters as part of the UKIDSS GPS survey (see Sect.\,\ref{observations}
for details), and by Spitzer with the IRAC camera at 3.6, 4.5, 5.8, and
8.0\,\mic\ as part of the GLIMPSE survey. Specifically we are mapping a region
that covers 10\deg\,$< l <$\,65\deg\ and $-$1.3\deg\,$< b <$\,+1.3\deg\ (left
panel of Fig.\,\ref{area}). This field includes GMCs associated with the distant
Scutum and 4\,kpc spiral arms of the Galaxy that comprise the northern portion
of the Molecular Ring (Dame et al. \cite{2001ApJ...547..792D}). The region also
includes many well-known massive star forming regions (W\,33, W\,51, etc.),
supernova remnants (W\,44, W\,49B,  etc. - sites of triggered star formation)
and galactic clusters (e.g. M\,16, M\,17, etc.), as well as a large number of
clouds within 1\,kpc of the Sun, with radial velocities $\le$20\,km/s (Dame \&
Thaddeus \cite{1985ApJ...297..751D}). The latitude range also includes the vast
majority of the molecular and atomic gas. All of the northern massive star
forming GMCs with masses of 10$^4$ to 10$^6$\,M$_\odot$ are covered by the
longitude range of our survey (Dame \& Thaddeus \cite{1985ApJ...297..751D}). It
includes the Galactic Ring at a galactocentric radius of 5\,kpc, where star
formation is particularly intense. Indeed, the scale-height for OB stars in
our Galaxy is of the order of 30\,--\,50\,pc (Reed \cite{2000AJ....120..314R},
Elias et al. \cite{2006AJ....132.1052E}). This is a clear indication that much
of the massive star formation occurs within the latitude range of our survey.

An unbiased survey was preferred over a survey which targets specific regions
based on, e.g. low-resolution CO maps or galactic radio surveys, for a number of
reasons. An unbiased survey, by its very nature, provides more reliable
statistics for a sizeable portion of the Galactic Plane. Also, the UWISH2 survey
not only addresses star formation, but also evolved stars; the latter will not
necessarily be associated with massive molecular clouds or bright radio sources.
Thirdly, given the abundance of high and low mass cores along the Galactic
Plane, we assumed that most regions in the survey would contain emission-line
sources. This assumption was to some extent based on preliminary WFCAM JHK and
\htwo\ observations of a region that included the well-known high-mass star
forming regions DR\,21 and W\,75\,N (Davis et al. \cite{2007MNRAS.374...29D};
Kumar et al. \cite{2007MNRAS.374...54K}). In this area ($l$\,=\,81.7\deg,
$b$\,=\,+0.5\deg; d\,=\,2-3\,kpc), jets, outflows, PDRs and young clusters were
identified in separate regions: in the main DR\,21 and W\,75\,N high mass cores,
but also along the western periphery of the W\,75\,N core (100\arcsec\ west of
W\,75\,N) and towards a newly-discovered low mass cloud, dubbed L\,906\,E,
200\arcsec\ west of DR\,21. These latter regions would not necessarily show up
as major features in low resolution dust continuum, radio or CO surveys, and
therefore could be missed in a pointed survey of GMCs.

Of course, there have been numerous other infrared imaging studies of low and
high mass star forming regions which reveal an abundance of \htwo\ line-emission
features associated with jets and outflows from YSOs (e.g. Davis \& Eisl\"offel
\cite{1995A&A...300..851D}; Miralles et al. \cite{1997ApJ...488..749M}; Hodapp
\& Davis \cite{2002ApJ...575..291H}; Walawender, Reipurth \& Bally
\cite{2009AJ....137.3254W}; Ginsburg et al. \cite{2009ApJ...707..310G};
Varricatt et al. \cite{2010MNRAS.404..661V}; Buckle et al.
\cite{2011MNRAS.inpress.B}). The GLIMPSE survey itself also hints at the
existence of many extended line-emission objects at locations along the Galactic
Plane. Dubbed Extended Green Objects (EGOs), the colour of these features may
result from enhanced line emission in the 4.5\,\mic\ Spitzer-IRAC band
(Cyganowski et al. \cite{2008AJ....136.2391}, Stecklum et al.
\cite{2009pjc..book..619S}).  These are often regarded as pure-rotational \htwo\
which in these extended objects may be associated with outflows from young
stars. However, there is growing evidence that the observed emission at least in
some objects can be alternatively attributed to scattered continuum emission
(e.g. Qiu et al. \cite{2008ApJ...685.1005Q}; De Buizer \& Vacca
\cite{2010AJ....140..196D}; Chen et al. \cite{2010ApJ.submitted.C}). Follow-up
observations at higher resolution, like those presented in this paper, are
needed to reveal the true nature of these objects.

\section{Observations}
\label{observations}

The narrow-band images that constitute UWISH2 are being obtained with the
Wide-Field Camera (WFCAM, Casali et al. \cite{2007A&A...467..777C}) at the
United Kingdom Infrared Telescope (UKIRT). The camera uses four Rockwell
Hawaii-II (HgCdTe 2048\,$\times$\,2048) arrays. The gaps between the four arrays
are equivalent to 94\,\% of the width of an array. The arrays have a pixel scale
of 0.4\arcsec; a contiguous area covering 0.75 square degrees on the sky (a
WFCAM tile) is imaged by observing at four discrete positions. In order to
correct for image artifacts, bad pixels, and to fully sample the point spread
function, a 2\,$\times$\,2 micro-stepping pattern is repeated at three jitter
positions. The jitter positions are offset by 6.4\arcsec\ (specifically
0\arcsec,0\arcsec; 6.4\arcsec,0\arcsec; 6.4\arcsec,6.4\arcsec); at each jitter
position we microstep with offsets of $\pm$\,1.4\arcsec\ (N+1/2 pixels on the IR
arrays and an integral number of autoguider pixels). This results in a pixel
scale of 0.2\arcsec\ in the final stacked images. An exposure time of 60\,s is
employed; the total per-pixel integration time is thus 720\,sec. The images are
acquired through a narrow-band filter ($\Delta\lambda =$\,0.021\,\mic ) centred
on the 1-0\,S(1) line of molecular hydrogen at 2.122\,$\mu$m.

All WFCAM data are reduced by the Cambridge Astronomical Survey Unit (CASU) and
are distributed through a dedicated archive hosted by the Wide Field Astronomy
Unit (WFAU) in Edinburgh, U.K.. The CASU reduction steps are described in detail
by Dye et al. \cite{2006MNRAS.372.1227D}; astrometric and photometric
calibrations (Hodgkin et al. \cite{2009MNRAS.394..675H}) are achieved using
2MASS (Skrutskie et al. \cite{2006AJ....131.1163S}, Dye et al.
\cite{2006MNRAS.372.1227D}; Hewett et al. \cite{2006MNRAS.367..454H}). Data are
then downloaded in bulk from WFAU and are made available (initially only to
members of the UWISH2 consortium) via a dedicated
web-site\footnote{http://astro.kent.ac.uk/uwish2/}. However, all data are
available directly from WFAU, and are made publicly available without
restriction 18\,months after they are acquired. The same applies to data
access via our dedicated web-site.

In order to continuum subtract our narrow band \htwo\ images we use the
K-band data obtained as part of the UKIDSS GPS (Lucas et al.
\cite{2008MNRAS.391..136L}). The \htwo\ and K-band images are aligned using the
astrometric calibration parameters stored in the file headers and scaled to the
same size. The flux scaling between \htwo\ and K-band is done for each star
individually. We then determine a map of the median flux scaling factor with a
resolution of 1\arcmin. This is required since the scaling factor between \htwo\
and K-band changes for regions with high extinction. To subtract the continuum
as accurately as possible, i.e. getting the least amount of residuals for stars
we also apply a Gaussian smoothing to the image with the better seeing to ensure
the same full width half maximum in both images before continuum subtraction.
Naturally, this process leads to 'negative bowl' effects, which can be used to
identify K-band reflection nebulae. The continuum subtracted \htwo\ images will
be made available on our dedicated web-site. Please note, that we have used
these images only for the identification of line-emission sources. They should
not be used for photometry.

To facilitate and simplify the continuum subtraction, we have tiled our survey
field in exactly the same way as the GPS. Due to the inclination of the Galactic
to the Equatorial Plane, at least four tiles are required to ensure that all
regions with $|b| <$\,1\deg\ are completely covered. Thus, our survey (once
completed) will cover a somewhat larger area, with $|b| \le$\,1.35\deg; in some
regions we will reach as far as $|b| =$\,1.7\deg\ or 1.8\deg. Hence, an area of
approximately 172.5\,square degrees in total will be imaged.

\section{Results}

\subsection{Survey Completeness}

The survey was originally awarded 222\,hrs of observing time spread over four
semesters (2009B, 2010A, 2010B and 2011A). Survey observations began on 28 July
2009 and progressed rapidly. As of September 2010 we have obtained data for 141
tiles. To image the entire envisaged survey area we have prepared 224 tile
observations. Hence, the survey is currently 62.9\,\% complete. For the first
three semesters we have achieved an efficiency of 83.9\,\%. The time lost
(16.1\,\% of the total) can be attributed to technical difficulties and a run of
bad weather in semester 2009B. The efficiency of the observations in 2010A and
2010B has been very close to 100\,\%.

The prospects for completion of the survey are somewhat unclear, owing to the
uncertain future of UKIRT. Hence, there will be no scheduled observations
for our project in  Semester 2011A. However, members of the survey team are
currently campaigning to use this project as bad-weather filler for the UKIDSS
surveys. Furthermore, we will obtain more data next summer via UH time access to
the telescope.

The area covered so far is displayed in the right hand panel of
Fig.\,\ref{area}.  Our initial goal was to observe a contiguous area at low
galactic longitudes (regions with low and negative Declinations), focusing also
on the Galactic mid-Plane. However, the inner galaxy sets an hour or so before
regions at higher $l$, so some tiles have been observed further out along the
Galactic Plane, and at higher latitudes. Even so, we have managed to observe a
contiguous area that covers $l =$\,10\deg\ -- 28\deg\ across the entire width
of the survey. Moreover, we have covered the entire Galactic mid-Plane ($|b|
<$\,0.5\deg) between $l > $\,10\deg\ and $l <$\,37\deg. Currently, out of the
completed tiles, about two thirds (63.4\,\%) are in the Galactic mid-Plane. The
region with the least amount of data taken so far is 37\deg\,$< l < $\,58\deg.
There, only about one quarter of the area has been observed. The remaining area
up to $l = $\,65\deg\ is approximately two-thirds complete.

\subsection{Data Quality}

\begin{figure}
\includegraphics[width=8.5cm]{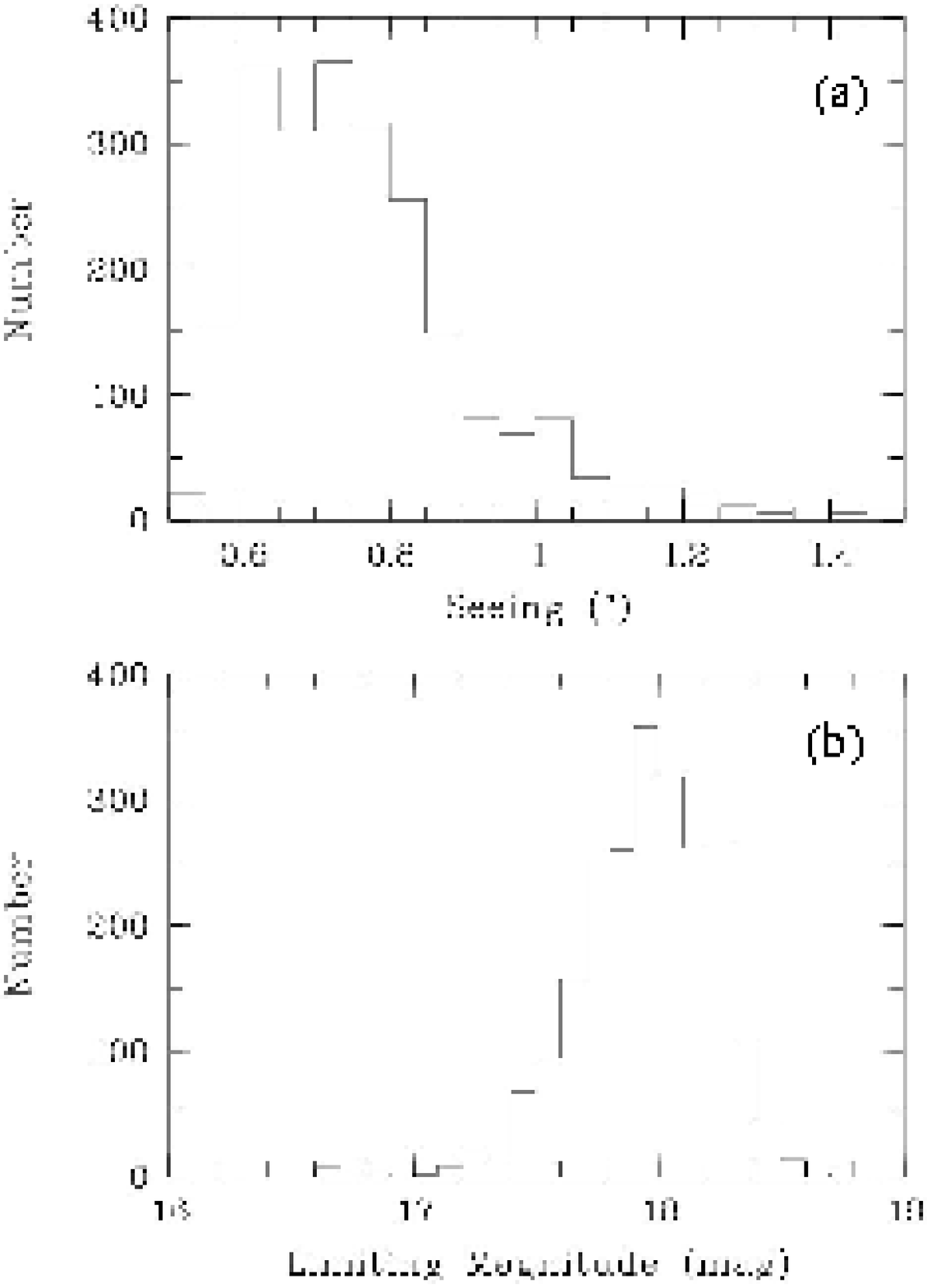}
\caption[]{\label{stats} Histograms showing seeing and limiting magnitude
(5\,$\sigma$ detection limit) statistics, measured from the 141 tiles observed
prior to 21 September 2010.}
\end{figure}

\begin{figure} 
\centering 
\includegraphics[height=8.5cm, angle=-90]{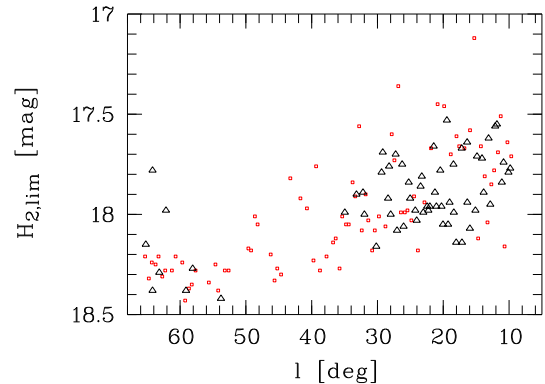} 
\caption[]{\label{long} A plot showing how the limiting magnitude (5\,$\sigma$
detection limit) measured in our \htwo\ images varies with Galactic Longitude
$l$. Red squares indicate tiles less than 0.5\deg\ from the Galactic Plane,
black triangles are tiles further than 0.5\deg\ from the Plane. The two points
at $l$\,$>$\,60\deg\ and limiting \htwo\ magnitute of less than 18\,mag are
taken in bad weather conditions.}
\end{figure}

Since the main goal of the survey was to obtain a catalogue of all \htwo\
emission line features, most of which are expected to be extended, the original
seeing limitations set for the UWISH2 survey were quite relaxed (less than
1.5\arcsec). However, this goal has been significantly surpassed for the vast
majority of our observed fields. The seeing statistics measured from the 141
tiles observed to date are displayed as a histogram in the top panel of
Fig.\,\ref{stats}. As one can see, the majority of the data have a seeing
between 0.6\arcsec\ and 0.8\arcsec\ in the co-added frames (the median seeing is
0.73\arcsec). Indeed 90\,\% of the images have a seeing better than 1.0\arcsec.
In some frames the seeing is as low as 0.51\arcsec, and only one image has a
measured seeing above the original project limit of 1.5\arcsec. 

The astrometric solution of our images has the same precision as the UKIDSS
GPS data, i.e. 0.09\arcsec\ (Lucas et al. \cite{2008MNRAS.391..136L}). Note that
the seeing might introduce variations to this number. We refrain from evaluating
the accuracy of the position determination of \htwo\ features, since this
depends strongly on the brightness, the signal to noise, and the 'shape' of the
features. It is hence impossible to derive a general value for the position
accuracy of \htwo-knots.

The limiting magnitude (5\,$\sigma$ detection limit) in the \htwo\ narrow band
filter ranges mostly between 17.6\,mag and 18.3\,mag (bottom panel in
Fig.\,\ref{stats}). The limiting magnitude in each image is generally not a
result of the sky quality (the vast majority of the data was taken under
photometric conditions), but rather is caused by crowding along the Galactic
Plane. This is nicely illustrated by the apparent correlation between the
limiting magnitude in the images and their position (longitude) along the
Galactic Plane (see Fig.\,\ref{long}). One can see in this figure that for
fields more than about 40\deg\ or 50\deg\ from the Galactic Centre, the limiting
factor is indeed the integration time, i.e. we reach a point source limit of
18.3\,mag. For fields closer to the Galactic Centre a linear correlation between
$l$ and the limiting magnitude is seen, caused by the increased crowding
(confusion noise) in the images at these positions.

For the detection of \htwo\ emission, which is usually extended, the surface
brightness detection limit is of interest. We estimate a typical {\it rms} noise
level in our \htwo\ images of 
3.4\,$\cdot$\,10$^{-19}$\,W\,m$^{-2}$\,arcsec$^{-2}$. We hence note that a
source with a uniform surface brightness of
10$^{-18}$\,W\,m$^{-2}$\,arcsec$^{-2}$ will be detected at a 3\,$\sigma$ level
in our unbinned images (0.2\arcsec\ pixel scale). Assuming the \htwo\ features
are extended over several arcseconds, we estimate a 3\,$\sigma$ detection limit
after re-binning our images to e.g. 1.2\arcsec\ (a typical resolution in
GLIMPSE) of 1.7$\cdot$\,10$^{-19}$\,W\,m$^{-2}$\,arcsec$^{-2}$. Using the
GLIMPSE 3\,$\sigma$ point source detection limit spread over the PSF
(3.4\,MJy\,str$^{-1}$, Churchwell et al. \cite{2009PASP..121..213C}) and a
conversion from \htwo\ flux at 4.5\,$\mu$m into the 1-0\,S(1) flux for a range
of typical conditions (Takami et al. \cite{2010ApJ...720..155T}) one finds that
the GLIMPSE 3\,$\sigma$ detection limit corresponds to
5-40\,$\cdot$\,10$^{-17}$\,W\,m$^{-2}$\,arcsec$^{-2}$ in the 1-0\,S(1) line of
molecular hydrogen. Our survey is hence a factor of 300 to 2000 better than the
corresponding GLIMPSE detections of \htwo\ in the 4.5\,$\mu$m filter. Thus, only
for regions with a K-band extinction in excess of 6\,mag, might GLIMPSE detect
4.5\,$\mu$m \htwo\ emission not detectable in our data. 

\section{Discussion}

\subsection{Star Formation}

\begin{figure*} 
\includegraphics[width=12cm]{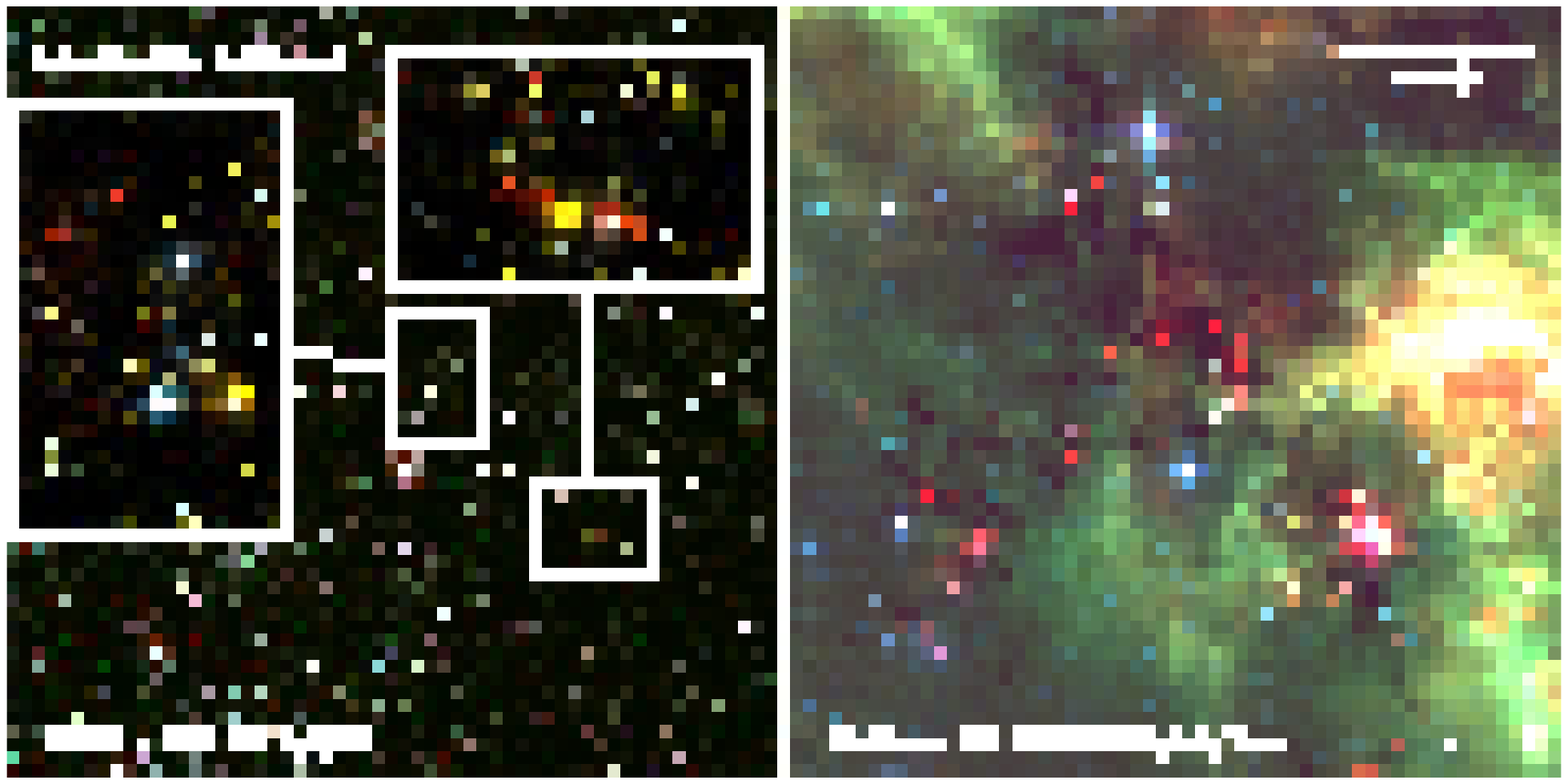}
\caption[]{\label{irdc} A comparison of UKIRT/WFCAM (UWISH2+GPS) imaging and
Spitzer/IRAC+MIPS (GLIMPSE+MIPSGAL) imaging of an IRDC at $l \sim $53.2\deg.
\htwo\ line-emission features, presumably associated with outflows, appear as
red features in the WFCAM data; these are largely undetected in the Spitzer
image, which instead illustrates the bright dust emission associated with this
region, and the chain of reddened sources associated with the IRDC. The IRDC
itself appears as a region devoid of stars in the WFCAM image, and is seen in
absorption against the dust emission in the Spitzer image. The yellowish
extended features in the left panel are red reflection nebulae, generally
coinciding with bright Spitzer sources. North is up and East to the left in
both images.}   
\end{figure*}

Collectively, jets and outflows can be used as sign-posts of star formation
(e.g. Bally et al. \cite{1995ApJ...454..345B}; Eisl\"offel
\cite{2000A&A...354..236E}; Froebrich \& Scholz \cite{2003A&A...407..207F};
Hodapp \cite{2007AJ....134.2020H}; Hatchell et al. \cite{2007A&A...472..187H};
Davis et al. \cite{2009A&A...496..153D}; Walawender et al.
\cite{2009AJ....137.3254W}). An abundance of jets points to active accretion and
a young population; a paucity of flows, in a region where near- and mid-IR
photometry still point to a sizeable population of reddened sources, indicates a
more evolved region with a larger population of pre-main-sequence stars (T-Tauri
and Herbig Ae/Be stars). Indeed, one of the goals of this survey is to establish
whether the large number of outflows seen in a massive star forming region like
DR\,21/W\,75 (over 50 independent flows have been detected in a single WFCAM
tile -- Davis et al. \cite{2007MNRAS.374...29D}), is common in other massive
star forming regions. Since the dynamical age of a protostellar outflow is 10 to
100-times less than the turbulent lifetime of a GMC, an abundance of outflows in
each region would point to ongoing or multiple epochs of star formation, rather
than infrequent bursts of star formation in each GMC.

Since outflows are also a direct tracer of mass accretion and ejection they can
be used to estimate star formation efficiency, particularly in high mass star
forming regions where the efficiency is grossly affected by molecular cores and
existing massive young stars, which influence the environment via their hugely
energetic winds and intense UV fluxes. An unbiased survey like UWISH2 will map
star formation efficiency from region to region, providing a reliable map of
the distribution of dynamically active (showing signs of jets/outflows) star
forming regions along the Galactic Plane.

Individually, jets and outflows can be used to pin-point the locations of
protostars. Flows from Class\,0 sources, for example, are typically 10-times
brighter in \htwo\ emission than their Class\,I counterparts (Caratti o Garatti
et al. \cite{2006A&A...449.1077C}), while optically-visible Herbig-Haro jets
from Class\,II sources are extremely faint in \htwo\ emission, because of a lack
of ambient molecular gas. Outflows from massive young stars are, on the other
hand, rarely detected in the optical because of extinction. These flows are
usually only seen in \htwo\ emission (e.g. Varricatt et al.
\cite{2010MNRAS.404..661V}) and in millimeter-wave line maps and radio continuum
images (Arce et al. \cite{2007prpl.conf..245A}). However, the spatial resolution
of the longer wavelength observations is usually poorer. With outflow sources
identified, the mass of the individual sources, derived from supporting
multi-wavelength photometry (WFCAM-GPS, Spitzer, JCMT-GPS, VLA, etc.) and thus
mass/luminosity estimates, can be measured. This sort of analysis will help to
address the question of whether the most massive stars, which may not form
through disk accretion, can generate collimated outflows. Because \htwo\ flows
are driven by the youngest sources, \htwo\ observations can also break the
protostar/T-Tauri star ambiguity in Spitzer-IRAC colour-colour analysis, where
inclination effects grossly affect mid-IR colours and thereby hinder individual
source classifications (Class\,0/I or Class\,II; e.g. Allen et al.
\cite{2004ApJS..154..363A}).

Jets are also powerful tracers of infall history, since jet parameters correlate
closely with mass infall rates and accretion luminosities (Beck
\cite{2007AJ....133.1673B}, Antoniucci et al. \cite{2008A&A...479..503A}). Tight
clustering, interactions between protostellar neighbours, and particularly
photo-evaporation and ablation of protostellar disks, can inhibit accretion and
thereby switch off the engine that drives an outflow (they can also trigger
accretion and cause FU-Ori type outbursts). This will be particularly important
in massive star forming regions, where young stars form in clusters and where
massive stars influence their lower-mass neighbours through gravitational,
radiative and mechanical (outflow) interactions. Estimates of the frequency of
jet activity in clustered environments, provided by UWISH2, will lead to an
assessment of the degree to which interaction inhibits accretion.

Statistical studies of jets can also shed light on the dynamics of cloud
collapse and star formation in GMCs. Are outflows randomly orientated, or are
they aligned perpendicular to cloud filaments though parallel to magnetic field
lines (Eisl\"offel et al. \cite{1994AJ....108.1042E}, Banerjee \& Pudritz
\cite{2006ApJ...641..949B})? Existing observations yield contrasting results
(e.g. Anathpindika \& Whitworth \cite{2008A&A...487..605A}; Davis et al.
\cite{2009A&A...496..153D}); clearly, a large, statistically-significant sample
of flows, with complementary observations of cloud and magnetic field
structures, is required to resolve this issue. The structure of the underlying
magnetic field is intricately linked to the fragmentation process of filamentary
clouds (e.g. Fiege \& Pudritz \cite{2000MNRAS.311..105F}, Falgarone et al.
\cite{2001ApJ...555..178F}), and consequently to outflow orientation. The
outflow sample obtained by our survey will be an outstanding and unique
opportunity to perform this analysis. We expect that the majority of outflow
sources will be at distances of 1\,--\,3\,kpc. The Galactic Plane magnetic field
structure within 2 or 3\,kpc is revealed by stellar polarization surveys (e.g.
Heiles \cite{2000AJ....119..923H}) and out to larger distances by other
techniques such as Faraday rotation of pulsars and extra-galactic radio sources
(e.g. Han \cite{2009IAUS..259..455H}). Recently the relation of outflow
direction and magnetic field structure has been investigated in individual
clouds (e.g. in DR\,21 by Kirby \cite{2009ApJ...694.1056K}). When such magnetic
field studies become available for other regions, the UWISH2 data will be
helpful to support or negate the relative orientation of outflows with magnetic
fields, thus constraining an important aspect of star formation.

There are of course many other outflow parameters that can be measured in a
large infrared study: What fraction are collimated, and what fraction are
parsec-scale in length? Does the mean flow length correlate with the median age
of the embedded population, derived from colour analysis or the mean mid-IR
spectral index measured from Spitzer-IRAC photometry? Are outflows from massive
stars better aligned (e.g. orthogonal to cloud filaments and/or large-scale
magnetic fields) than their less powerful, lower-mass counterparts? And can
outflows account for the turbulent motions in GMCs; are they energetic enough
and sufficient in numbers, and what is the relative role of massive YSO flows
and low-mass YSO outflows? Our goal is to address many of these issues in
future analysis of UWISH2 data.

\begin{figure}
\centering 
\includegraphics[width=8.5cm]{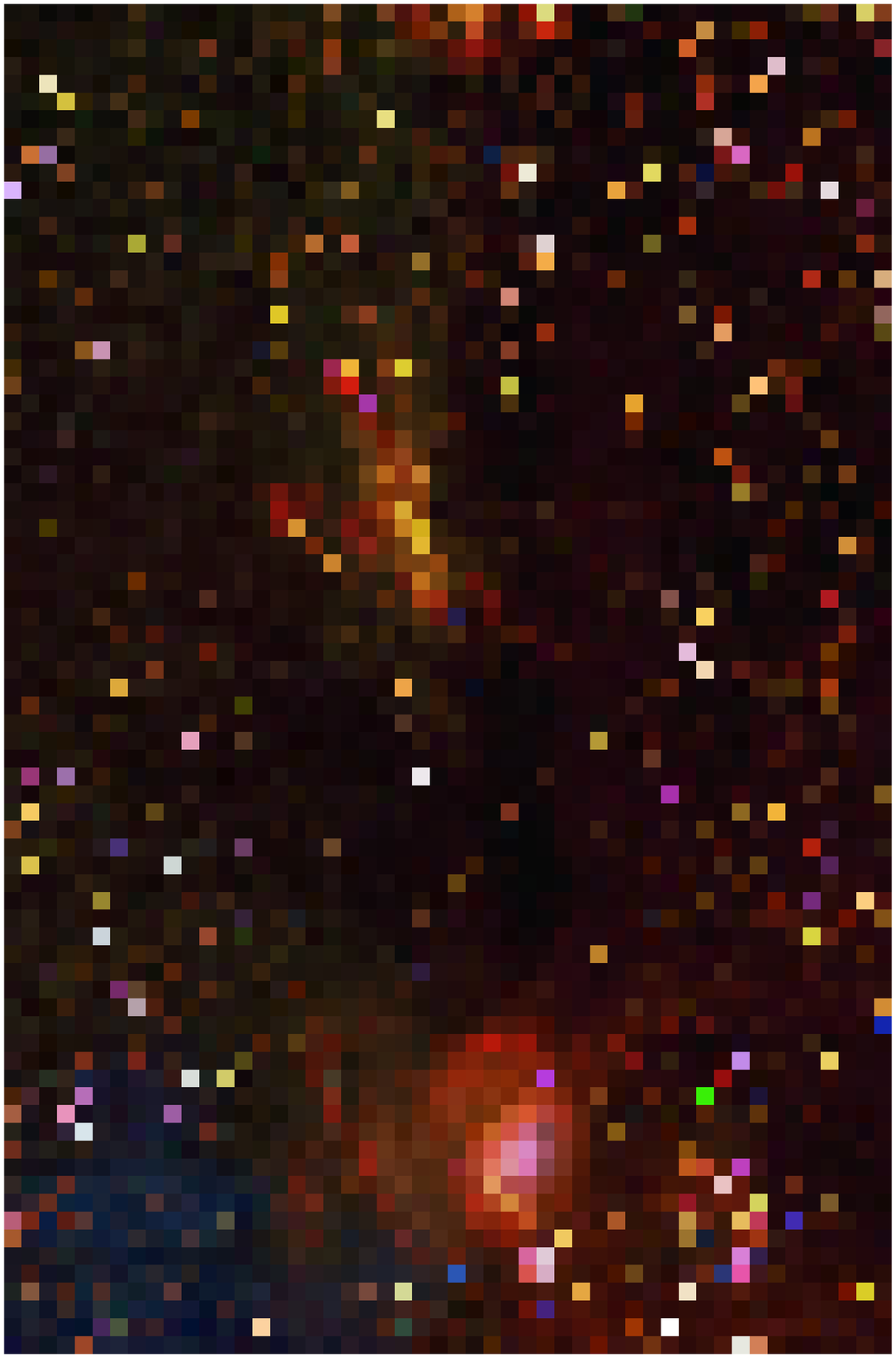}
\caption{\label{g35} A JK\htwo\ colour-composite image showing the
remarkable  massive star forming region G\,35.2\,N (just north of the centre),
the young cluster Mercer\,14 (in the south-western corner) and an H{\small II}
region (just east of the cluster). The green dots indicate the outflows
identified to the North-East of the young cluster Mercer\,14. The image is about
5\arcmin\,$\times$\,7\arcmin\ in size. North is up and East to the left in the
image.} 
\end{figure}

Examples of two rather different star forming regions are shown in
Figs.\,\ref{irdc} and \ref{g35}. The GLIMPSE+MIPSGAL Spitzer data presented in
the former reveal a remarkable Infra-Red Dark Cloud (IRDC), along which there
appears to be a number of reddened mid-IR sources. IRDCs seen in silhouette
against the bright Galactic background in the mid-IR are believed to be the
precursors to massive stars and star clusters, but individual IRDCs show diverse
star forming activities within them (Henning et al. \cite{2010A&A...518L..95H}).
The IRDC at $l \sim 53.2^\circ$ presented in Fig.\,\ref{irdc} is an intriguing
example of such clouds. It is associated with a long, filamentary CO cloud at
2\,kpc distance that extends to the north-east, beyond the boundary of the
figure, in the Galactic Ring Survey data of $^{13}$CO\,$J$\,=\,1\,--\,0 emission
(Jackson et al. \cite{2006ApJS..163..145J}). Thus, the total extent of the IRDC
reaches $\sim 30$\,pc. This IRDC was partially identified as three separate,
arcmin-size clouds by Simon et al. \cite{2006ApJ...639..227S} in their catalog
of MSX IRDC candidates. One of these is associated with the most impressive
outflow in Fig.\,\ref{irdc} in the south-western area. Each cloud has an
estimated mass of 100\,--\,500\,$M_\odot$ and is composed of several compact
millimeter-continuum cores (Rathborne et al. \cite{2006ApJ...641..389R}, Simon
et al. \cite{2006ApJ...653.1325S} Butler \& Tan \cite{2009ApJ...696..484B}). In
Fig.\,\ref{irdc}, the complementary UWISH2+GPS data reveal ubiquitous outflows
along the IRDC. Detailed analysis of this field is reserved for a future paper
(Kim et al. in preparation). However, the fact that some of the reddened mid-IR
sources have  associated outflows while some do not indicates that the IRDC is a
site of active star formation with YSOs in various evolutionary stages.

Figure\,\ref{g35} shows the JK\htwo\ colour-composite of the area containing the
high-mass star forming region G\,35.2\,N, the young cluster Mercer\,14 and an
{H{\small II} region. The massive molecular outflow previously observed in CO
and radio by Birks et al. \cite{2006A&A...458..181B} and De Buizer
\cite{2006ApJ...642L..57D} is evident in bright \htwo\ line emission. The area
contains CO outflows (Dent et al. \cite{1985A&A...146..375D}) as well as H$_2$O,
methanol and OH masers (Caswell et al. \cite{1995MNRAS.272...96C}; Hutawarakorn
\& Cohen \cite{1999MNRAS.303..845H}). Observations suggest that the outflows are
driven by more than one source and that the distance to the complex is about
2\,kpc (Brown et al. \cite{1982MNRAS.201..121B}). $^{\rm 13}$CO data, taken as
part of the RMS-survey by Urquhart et al. \cite{2008A&A...487..253U} indicate a
distance of 2.3\,kpc. These authors also determine the bolometric luminosity of
the source to be about 1.7\,$\cdot$\,10$^4$\,L$_\odot$. 

Using 2.3\,kpc as the distance, we measure the length of both \htwo\ outflows to
be about 1.2\,pc. The \htwo\ flows lie on the same principle outflow axis as
observed in the radio observations. The bright bow-shaped features, in
particular in the north-eastern lobes, seem to form the terminal bow-shock of
the flow interacting with the surrounding medium. The blue and redshifted lobes
point to the north-east and south-west, respectively. This is in agreement with
the strength of the \htwo\ emission, which is higher for the blue-shifted part
of the outflow. We measure a combined, de-reddened (assuming $A_K$\,=\,2\,mag)
1-0\,S(1) luminosity of about 1\,L$_\odot$. Assuming typical shock conditions,
about 10\,\% of the entire \htwo\ emission will be in the 1-0\,S(1) line (Smith
\cite{1995A&A...296..789S}, Caratti o Garatti et al.
\cite{2006A&A...449.1077C}), but note that the cold component, traced by the
0\,--\,0 transitions might play a major role in the overall energy balance
(Smith \& Rosen \cite{2005MNRAS.357.1370S}, Caratti o Garatti et al.
\cite{2008A&A...485..137C}). Hence, the total \htwo\ luminosity is about
10\,L$_\odot$. This is in good agreement with the expected relation between
source bolometric luminosity and outflow \htwo\ luminosity for young Class\,0
and Class\,1 protostars (discussed e.g. in Caratti o Garatti et al.
\cite{2006A&A...449.1077C}). Together with IRAS\,20126+4104 (Caratti o Garatti
et al. \cite{2008A&A...485..137C}), this object is one of the few luminous young
objects where the outflow has been observed in such detail in \htwo\ emission.
Our survey will investigate many more of these massive outflows and investigate
the relation between \htwo\ emission and the outflow driving sources.

Presumably unrelated to G\,35.2\,N, outflow activity is observed just to the
north-east of the young cluster Mercer\,14. At least three outflows can be
identified in that region. To the east of the young cluster we identify an
{H{\small II} region, surrounded by most likely fluorescently excited \htwo\
emission as well as a continuum reflection nebula. The object is listed as a
planetary nebula (PN\,G035.1$-$00.7) but given the appearance, this is clearly a
mis-classification. A more detailed analysis of this region will be performed in
Ioannidis et al. (in preparation).

A preliminary analysis of UWISH2 data obtained in the first semester of
operations (2009B) has been conducted. Images covering 24 square degrees have
been examined. We identify about 350 candidate \htwo\ outflows and/or emission
line features (Molecular Hydrogen emission-line Objects - MHOs; Davis et al.
\cite{2010A&A...511A..24D}) distributed over about 50 star forming regions.
Based on this cursory analysis of the first semester data, we estimate that
$\sim$2400 outflows spread throughout $\sim$300 star forming regions may be
detectable in the UWISH2 data, a remarkable result -- almost trebling the number
of known MHOs (Davis et al. \cite{2010A&A...511A..24D}). We investigated the
already analysed data for any trends of outflow numbers per square degree with
galactic longitude or latitute. No significant trends could be found. Rather the
number of flows per square degree varies by more than an order of magnitude
between a few and up to 80, when averaged over one tile. Given the small
fraction (just about ten percent of the total survey area) of data
analysed so far we cannot draw any meaningful conclusions. We will have to await
the investigation of the entire dataset before a complete and conclusive picture
emerges.

\begin{figure*} 
\includegraphics[width=12cm]{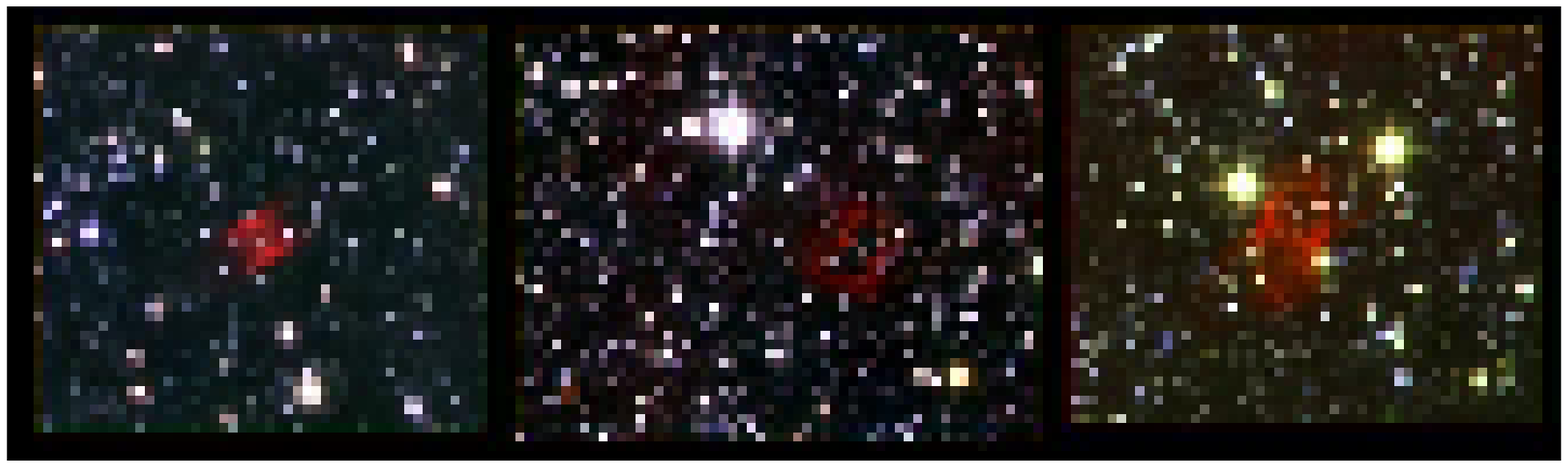}
\caption{\label{pn} JK\htwo\ Colour-composite images showing three of the newly
discovered planetary nebulae. The objects are UWISH2\,PN\,1, 4, 3. See
Table\,\ref{pnetab} for their positions and apparent sizes. North is up and East
to the left in all images.}
\end{figure*}

\subsection{H{\small II} Regions and Photon-Dominated Regions}

UWISH2 complements perfectly the CORNISH VLA survey at 5\,GHz and the IPHAS
H$\alpha$ survey, both of which are detecting not only evolved, extended
H{\small II} but also Ultra Compact H{\small II} (UCH{\small II}) regions by
virtue of their radio continuum and hydrogen recombination line emission. The
spatial resolution of WFCAM is well matched to the resolution of the VLA and
Spitzer surveys. Therefore, a statistical comparison can be made between radio
and molecular emission properties.


In the Spitzer-GLIMPSE data, Churchwell et al. \cite{2006ApJ...649..759C}
recently detected over 300 dusty bubbles surrounding late O or B-type stars.
Roughly 25\,\% coincide with known H{\small II} regions, while 13\,\% enclose
star clusters. Similar structures have been reported in ground-based \htwo\ (and
\brg) observations of O/B stars and associated PDRs in NGC\,7538 (Bloomer et al.
\cite{1998ApJ...506..727B}) and in WFCAM observations of DR\,21/W\,75 (Davis et
al. \cite{2007MNRAS.374...29D}; Kumar et al. \cite{2007MNRAS.374...54K}). A
spectacular example of a PDR is evident in the UWISH2 image in Fig.\,\ref{g35}.
Combining UWISH2, CORNISH and IPHAS observations will also allow us to establish
what fraction are fluorescently illuminated and what fraction are ionised. 
Thus, these data can be used to further examine the morphological and excitation
properties of these H{\small II} regions, along with their associated PDRs and
mid-IR bubbles. UWISH2 clearly offers sufficient resolution to distinguish
morphological features in many of these objects.  With (kinematic) distance
estimates, size scales may also be attributed to many of these objects, yielding
number distribution estimates and therefore lifetimes for \htwo\ -bright
H{\small II} regions.


Ground-based \htwo\ observations of PDRs may also allow further constraints to
be placed on current models of the material in PDRs. Allers et al.
\cite{2005ApJ...630..368A} have used \htwo\ observations of the Orion Bar PDR
(at a similar pixel resolution to that of WFCAM) in conjunction with Far-UV
information from the literature to identify a separation between the ionization
front and the \htwo\ emission peak, placing new constraints on both the FUV
attenuation cross-section and atomic heating rates used in PDR models. UWISH2
survey data, when used in conjunction with observations at other wavelengths,
will allow a whole range of PDR geometries, morphologies and radiation field
strengths to be studied in a similar manner.

\begin{table}
\caption{\label{pnetab} Positions and apparent diameters of the four newly
identified Planetary Nebulae. We list the diameter of the inner region without
detectable \htwo\ emission (D$_{\bf in}$) as well as the size of the maximum
extend of detectable emission (D$_{\bf out}$). In the cases where the object is
not symmetric, we list the range between the shortest and largest diameter.}
\centering
\renewcommand{\tabcolsep}{4pt}
\begin{tabular}{lrrrr}
Name & RA \,(J2000)& DEC\,(J2000) & D$_{\bf in}$\,[\arcsec] & D$_{\bf out}$\,[\arcsec] \\ \hline
UWISH2 PN 1 & 18:50:14.1 & +02:18:08 & 3.3  & 7.5   \\
UWISH2 PN 2 & 18:57:35.8 & +02:27:02 & 6.0  & 14-21 \\
UWISH2 PN 3 & 19:01:03.1 & +01:57:28 & 10.0 & 15-25 \\
UWISH2 PN 4 & 19:31:10.7 & +19:29:06 & 5-11 & 15-23 \\
\end{tabular}
\end{table}

\subsection{Post-AGB Stars and Planetary Nebulae}

Although widely distributed throughout the Milky Way, the post-AGB to PN phase
is relatively brief. Thus, only a few hundred are known in the Galaxy. Many
post-AGB stars and PN are bright \htwo\ emitters. Near-infrared spectroscopic
surveys have shown that objects in the early post-AGB phase can produce strong
\htwo\ emission as fast winds begin to blow and drive shocks into the
surrounding molecular envelope (e.g. Hrivnak, Kwok \& Geballe
\cite{1994ApJ...420..783H}; Garc\'{i}a-Hern\'{a}ndez et al.
\cite{2002A&A...387..955G}). As the central star evolves to hotter temperatures
the \htwo\ transitions are readily UV-pumped so that objects with B type central
stars (and hotter) show fluorescent \htwo\ emission (e.g. Davis et al.
\cite{2003MNRAS.344..262D}, Kelly \& Hrivnak \cite{2005ApJ...629.1040K}).

When combined with the UKIDSS-GPS, CORNISH 5\,GHz and IPHAS H$\alpha$ surveys,
UWISH2 will result in a more complete census of evolved sources within the
Galactic Plane. In particular, it will give a near extinction-free estimate of
PN numbers, revealing embedded young PN hidden by interstellar extinction.
Indeed, in our preliminary analysis of the 2009A data noted above, we have found
four new PN, three of which are shown in Fig.\,\ref{pn}. These objects were
selected based on the appearance of their \htwo\ line emission. We list the
positions and apparent sizes of the PN in Table\,\ref{pnetab}. The appearance of
the objects in \htwo\ emission ranges from compact circular symmetric objects to
bipolar and asymmetric. If we assume the objects have an average size of about
0.3\,pc (O'Dell \cite{1962ApJ...135..371O}), their distances range between
3\,kpc for the larger objects and about 8\,kpc for the compact UWISH2\,PN\,1.
This is a general argument and may not have any meaning in terms of these
specific objects as it depends on exactly the stage of evolution they are in.
Old PN will tend to be bigger (e.g. the median radius of a volume limited sample
within 1\,kpc is 0.6\,pc, which is dominated by old PN (Frew \& Parker
\cite{2010PASA...27..129F}). Young objects will be smaller and pre-PN
(protoplanetary nebulae) an order of magnitude smaller (e.g. 0.05\,pc).

\begin{figure*} 
\includegraphics[width=8.5cm]{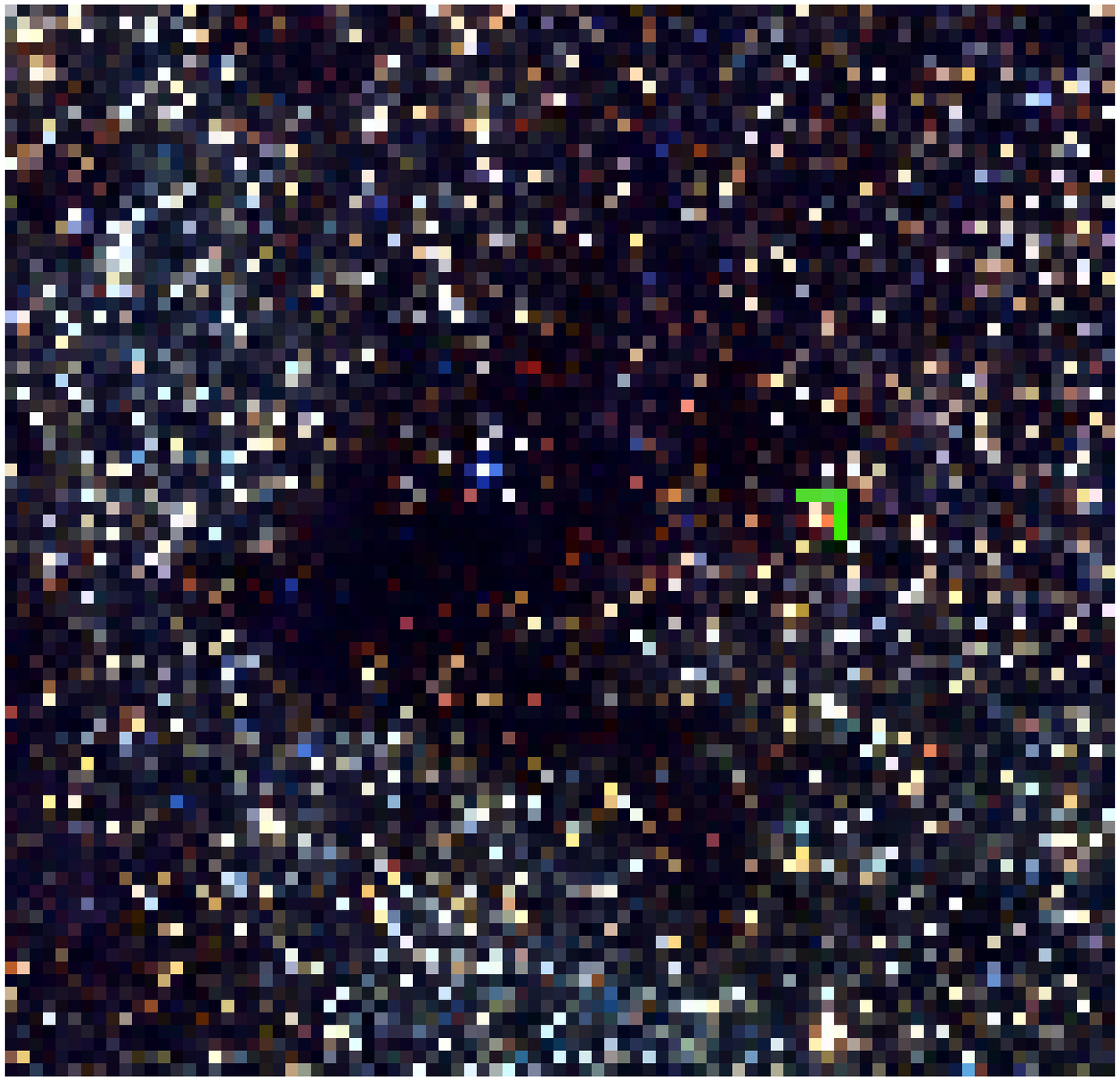} \hfill 
\includegraphics[width=8.5cm]{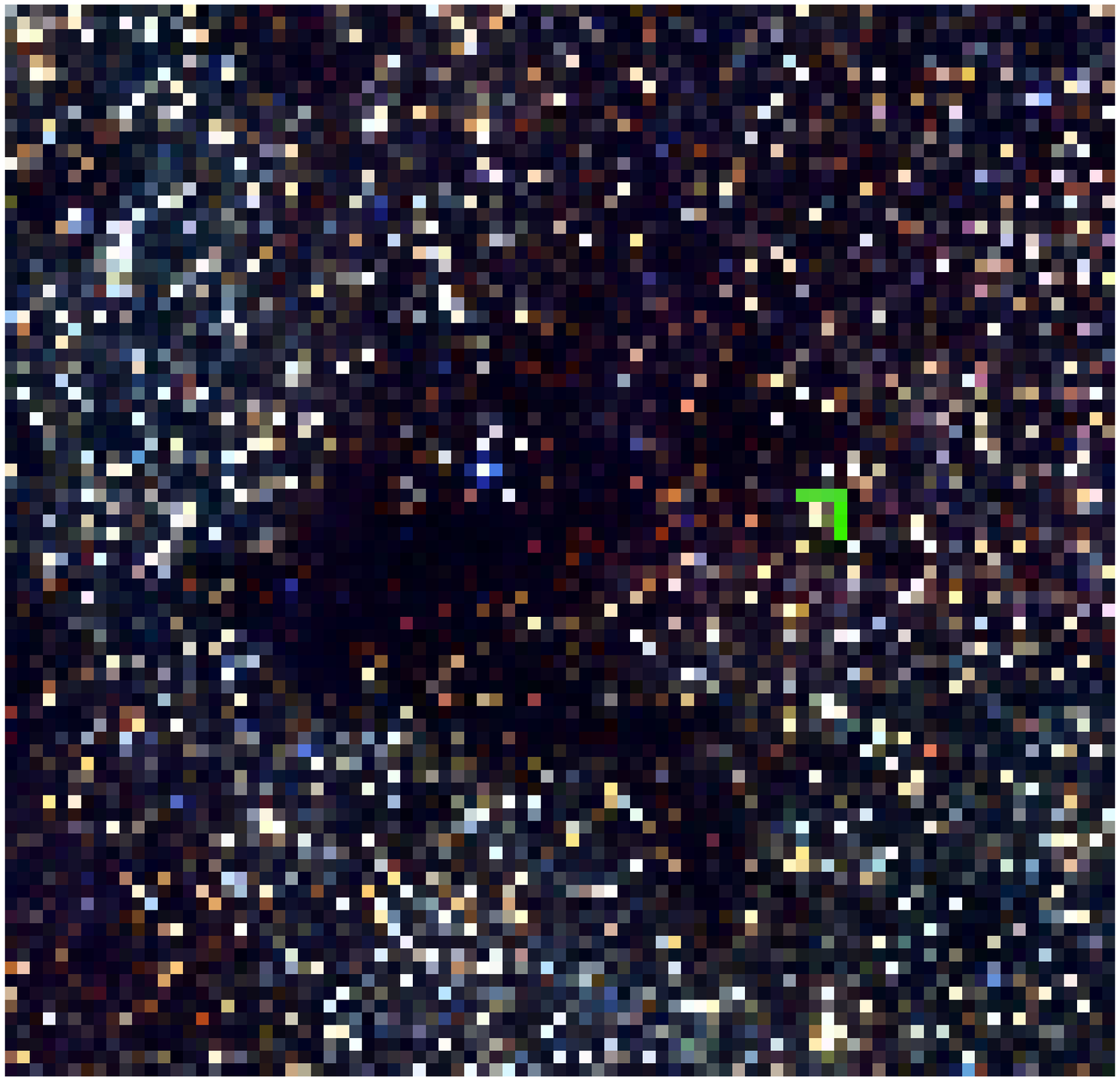}
\caption[]{\label{var} Colour-composite images showing the variable star 2MASS
J18072340-2025532 (marked by the green box), detected in one of our target
fields. The left panel is a JHK composite constructed from GPS data, while the
right panel is a JHH$_2$ composite using GPS and our data. The star clearly is
fainter in the UWISH2 images compared to the K-band UKIDSS GPS data obtained
about two years earlier. See text for more details. North is up and East
to the left in both images.}
\end{figure*}

The current number of identified PN in the Galaxy now stands at around 3000 and
is likely to rise as a result of ongoing surveys (e.g. Frew \& Parker,
\cite{2010PASA...27..129F}), including UWISH2. An accurate measure of the PN
content of the Galaxy is important in determining the relative importance of
the various routes to PN formation and the role of binary interaction (see de
Marco \cite{2009PASP..121..316D} for a review). It has been noted that \htwo\
-emitting PN tend to lie at low Galactic latitude, with a scale height of
150\,pc (Kastner et al. \cite{1996ApJ...462..777K}) and the same trend exists
for the pre-PN \htwo\ emitters, which tend to be bipolar (Kelly \& Hrivnak
\cite{2005ApJ...629.1040K}). A deep Galactic plane \htwo\ survey, such as
UWISH2, has the potential to uncover a significant population of evolved
objects and its contiguous areal coverage will provide important constraints on
their space density and any variation along the plane.

 %
 %

The UWISH2 survey also provides a unique study of PN morphologies (note the
considerable improvement in spatial resolution over Spitzer evident in
Fig.\,\ref{irdc}), particularly when complemented by IPHAS optical and CORNISH
radio survey data. Point-symmetric structure is seen in many post-AGB objects
and PN; this is thought to result from the interaction of a precessing jet with
the remnant AGB shell (Kwok \cite{2000eaa..bookE5200K}). In many ways the
physics of post-AGB evolution mimics that seen in pre-main-sequence objects,
something that could be investigated further with these simultaneous
observations of both classes of object. Observations of the near-IR \htwo\ and
CO bandhead emission in the post-AGB object IRAS\,18276-1431 show striking
similarities with emission features in Herbig Haro jet sources (Gledhill et al.
\cite{2010arXiv1009.5608G}). Our near-IR observations will certainly lead to a
more complete understanding of PN formation and shaping, and provide
observational constraints on (magneto) hydrodynamic and photo-ionization models
of PN formation and evolution.

Finally, to some extent UWISH2 may be useful in slecting transition objects,
specifically those that are too cool to produce radio or \brg\ emission, but
which produce shocked \htwo\ emission. The combined radio, \htwo\ and H$\alpha$
data will ultimately yield an evolutionary sequence for evolved stars based on
excitation, providing age determinations for each evolutionary stage from source
statistics.

\subsection{Variable and High Proper Motion Stars}

When combined with the GPS K-band data, UWISH2 is ideal for revealing long-term
variables (e.g. giants, FUors, EXors). Although the time between the GPS JHK and
UWISH2 \htwo\ observations is somewhat random, the bulk of the First Galactic
Quadrant was observed in the broad-band filters between May 2005 and July 2007,
about two to four years prior to the first \htwo\ observations were acquired in
July 2009. We also stress that the narrow-band images reach a similar depth to
point sources in the K-band (the GPS broad-band data reaches a modal depth of
K\,$\sim$\,18\,mag in uncrowded fields; Lucas et al.
\cite{2008MNRAS.391..136L}). Note that the GPS includes a 2$^{\rm nd}$ K band
epoch for which data collection is currently underway. Hence there will be three
epochs in total for much of the UWISH2 region. The broad and narrow-band data
can therefore be used to search for variable stars. 

We find that for most fields the conversion of \htwo\ into K-band magnitudes
required for this variability search, is straight forward and only minor offsets
and small colour terms need to be applied. In all cases using a $H-K$ colour
term leads to the conversion with the lowest $rms$ of the difference in flux
measurements of stars between epochs. However, the exact magnitude of the
offsets and colour terms varies with the observed field. This is due to known
issues with the photometric calibration. As noted in the WFCAM Science Archive
under "Known Issues", a small proportion of UKIDSS GPS fields with high
extinction throughout, currently have significant photometric calibration
errors. Sources in these fields are flagged with the "pperrbits" error flags in
the 7$^{\rm th}$ UKIDSS GPS data release. E.g. sources with photometric errors
in the K band have {\tt k$\_$1pperrbits} $\geq$\,131072, compared with {\tt
k$\_$1pperrbits} $<$\,256 for sources with a reliable calibration. This issue is
being investigated with the intention of correcting it for the 8$^{\rm th}$
UKIDSS GPS data release in early 2011.

We search for variable stars by converting the \htwo\ magnitudes into the
K-band, considering the colour of the objects. Then we search for objects that
change their brightness by more than 5\,$\sigma$ with respect to stars of
comparable magnitude. One example of a variable source detected in our survey
(2MASS J18072340-2025532) is shown in Fig.\,\ref{var}. This object has an IRAS
and MSX detection, is associated with an OH maser and listed as 'Star with
envelope of OH/IR type' in the
SIMBAD\footnote{http://simbad.u-strasbg.fr/simbad/} database. The source appears
much ``greener'' in our JHH$_2$ colour composite compared to the JHK composite
from the UKIDSS GPS (see Fig.\,\ref{var}), hinting at a dimming in the K-band
flux in the time between the observations. Indeed, as can be seen in
Table\,\ref{varstar}, the source shows a remarkable change in near infrared
brightness and colours over the last 11\,yrs. It changes its J-K colour from
about 6.5\,mag to about 3.2\,mag while brightening in the J-band and getting
fainter in the K-band.

Our examination of the 24 square degrees observed in 2009B revealed about 450
candidate variable stars, many of which will be variable giants or
FUor/EXor-type objects. Thus, throughout the entire UWISH2 survey area, we
expect to observe as many as 3000 variable sources spread along the Galactic
Plane. The threshold for detection of variability depends on the star's
brightness. We find that bright stars (K\,$<$\,15\,mag) that change their K-band
magnitude by more than 0.1\,mag between UWISH2 and UKIDSS GPS can be identified
at the 3\,$\sigma$ level. This increases up to about 0.7\,mag for stars at the
completeness limit of the survey. This should enable us to find not only
variable giants, FUORs and EXORS but also many of the young stars, and other
species as well.

Combining the UWISH2 and UKIDSS GPS surveys, with their sub-arcsecond
resolution, we are also well equipped to find stars with high proper motion
(HPM). These objects are also generally nearby, and hence a search for HPM stars
naturally contributes to our inventory of the immediate Solar neighbourhood.
Currently there is a lack of known nearby stars in the area of the Galactic
Plane, simply caused by the vast number of stars, crowding and the insufficient
spatial resolution of available surveys. The GPS has already been used to find
one of the nearest known brown dwarfs (UGPS\,0722-05, Lucas et al.
\cite{2010MNRAS.408L..56L}). The object was selected using its colours, but was
shown to have a proper motion of about 970\,mas\,yr$^{-1}$ and is at a distance
of 4.1\,pc. 

We cross-matched a fraction of our survey with the GPS K-band data to identify
detection limits for proper motion measurements. If one assumes a typical time
gap of four years between the observation, then the 3\,$\sigma$ detection limit
for PM measurements will be of the order of 100 to 200\,mas\,yr$^{-1}$. The
actual value will depend on the brightness of the stars, the local crowding and
the seeing conditions. Nevertheless, objects such as the above mentioned
UGPS\,0722-05 will be easily uncovered. Given our photometric detection limits
(see Fig.\ref{long}) of at least 17.5\,mag in the K-band, and the absolute
magnitudes of known brown dwarfs from Marocco et al. \cite{2010A&A...524A..38M}
we can determine the distances out to which a search for HPM stars can be
performed. We find that we can in principle search out to 80, 50, 20, and 5\,pc
for objects of spectral type T0, T5, T7, and T9, respectively. Of course the
actual limits will depend on the local crowding, blending, the corresponding
K-band photometric quality, and the actual proper motion. Note that away from
the Galactic Centre with a detection limit of 18.3\,mag in the K-band, the
distance out to which these objects are detectable increases by 45\,\%, or in
other words a three times larger volume can be searched.

\begin{table}
\caption{\label{varstar} Near infrared data of 2MASS J18072340-2025532. All
magnitudes are in the survey specific photometric systems. $^*$The K-band flux
of the star in the UWISH2 data has been estimated from the \htwo\ filter
magnitude by applying the mean offset between UWISH2 \htwo\ and GPS K-band
magnitudes for all stars in the field.}
\centering
\renewcommand{\tabcolsep}{4pt}
\begin{tabular}{llrrrr}
Survey & Date & J\,$[mag]$ & H\,$[mag]$ & K\,$[mag]$ & J-K\,$[mag]$ \\ \hline
DENIS & 1998/08/13 & 14.132 &  ---   & 7.734 & 6.398 \\
2MASS & 1999/07/07 & 14.797 & 10.545 & 8.026 & 6.771 \\
UKIDSS& 2007/05/09 & 12.353 & 10.440 & 9.070 & 3.283 \\
UWISH2& 2009/07/30 &  ---   &  ---   & 9.94$^*$ &  ---  \\
\end{tabular}
\end{table}

\subsection{Supernovae Remnants}

Supernova remnants (SNR) are another category of astronomical sources where the
\htwo\ line is of considerable use. \htwo\ lines are usually emitted by shocked
interstellar or circumstellar molecular gas swept-up by SNR shocks. But
sometimes the emission could originate from \htwo\ gas in pre-shock region
heated by a precursor or, in pulsar wind nebulae, from \htwo\ gas heated by
synchrotron emission and/or cosmic rays (Graham et al.
\cite{1990ApJ...352..172G,1991AJ....101..175G}). These \htwo\ lines can be used
to explore the physical properties and the nature of SNRs.

Among the known 274 Galactic SNRs in the Catalogue by Green
\cite{2009BASI...37...45G}, about 25\,\% are included in the UWISH2 area.
According to our preliminary analysis, at least seven of these SNRs show
associated \htwo\ emission. Figure\,\ref{w44} shows a JK\htwo\ colour-composite
image of the SNR W\,44 (G\,34.7$-$0.4), one of the most \htwo-prominent SNRs in 
the survey area. W\,44 is a middle-aged ($\sim 2\times 10^4$\,yrs) SNR
interacting with a molecular cloud at 3\,kpc and the \htwo\ emission had been 
detected by Reach et al. \cite{2005ApJ...618..297R}. However, this is the first
time that we see the entire remnant in \htwo\ emission. An extensive, organized
system of thin and knotty \htwo\ filaments is visible with the bright ones
delineating the SNR boundary. The \htwo\ image shows that the remnant has a
morphology that is extended along the northwest-southeast direction, which is
somewhat different from its morphology in radio emission (Castelletti et al.
\cite{2007A&A...471..537C}). This might be due to the non-uniform structure of
the ambient medium. Meanwhile, the \htwo\ morphology is quite similar to that
seen in the Spitzer IRAC 4.5\,$\mu$m band (Lee \cite{2005JKAS...38..385L}),
which implies that the latter is dominated  by \htwo\ lines in this band.
Detailed analysis of the UWISH2  data will reveal more SNRs with associated
\htwo\ emission, which  will help us to understand their physical environment
and evolution.

\begin{figure*}
\includegraphics[width=8.5cm]{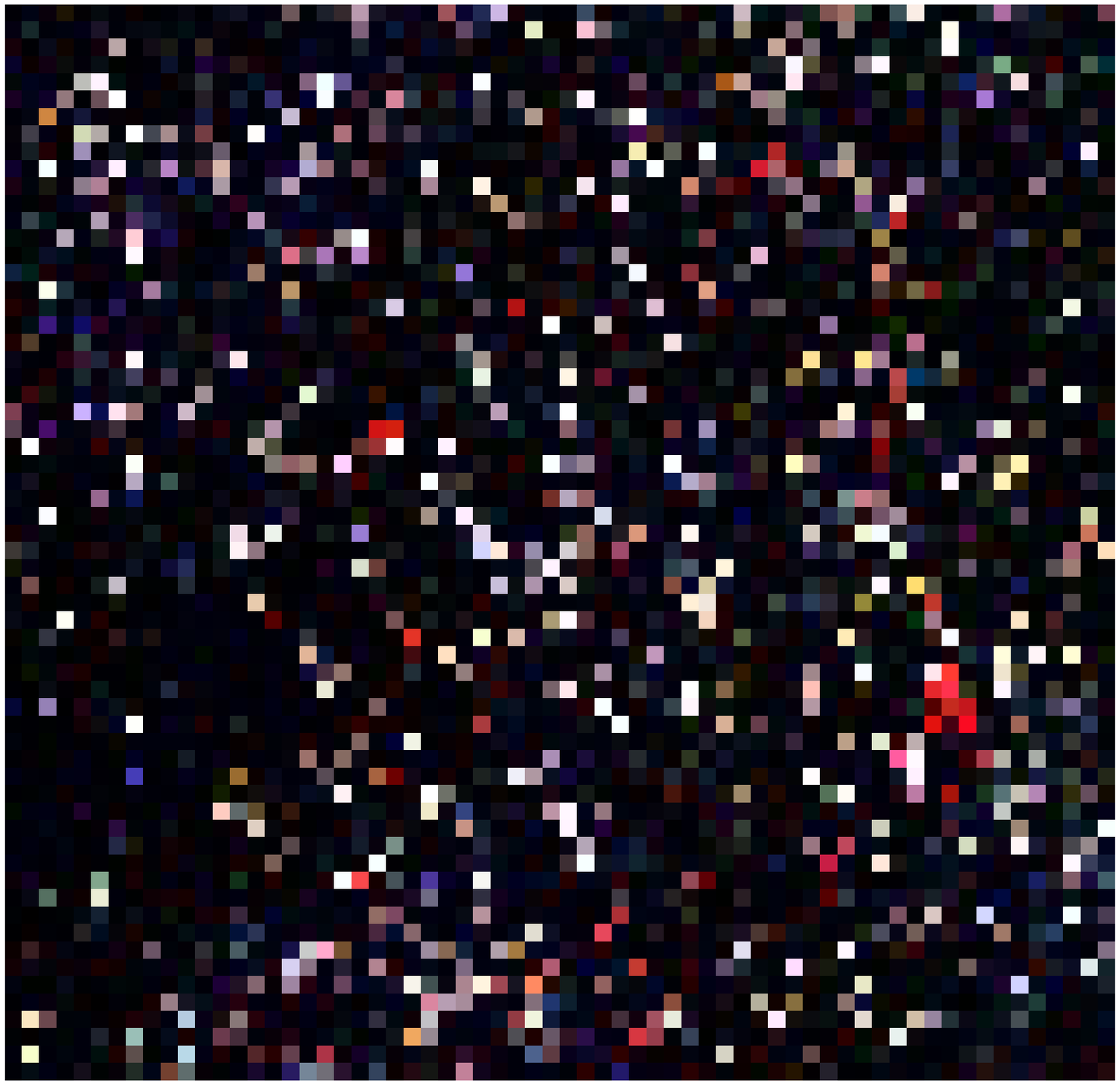} \hfill
\includegraphics[width=8.5cm]{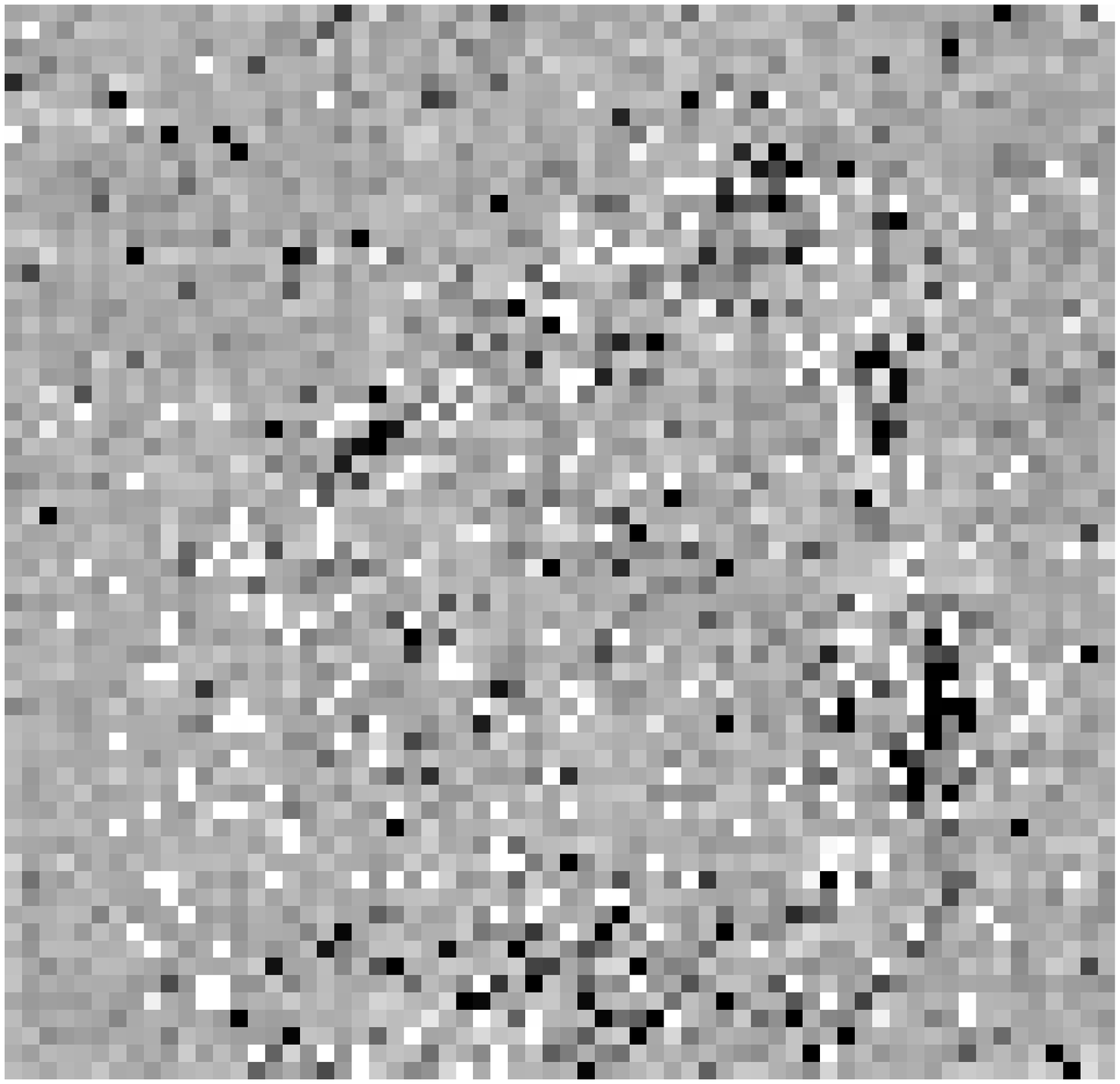}
\caption{\label{w44} {\bf Left:} A JK\htwo\ colour-composite image showing the
shocked \htwo\ filaments in the SNR W\,44. {\bf Right:} A \htwo-K gray scale
image of the same area. The white contours show the VLA 1.4\,GHz brightness
distribution from Giacani et al. (1997). North is up and East to the left in
both images.}    
\end{figure*}

\subsection{Evolved Massive Stars}

A fundamental key to understanding the evolution of massive stars is their mass
loss (e.g. Maeder \& Meynet \cite{2000A&A...361..159M}) and observations of
circumstellar nebulae around evolved massive stars (Wolf-Rayet stars, luminous
blue variables (LBVs), etc.) could provide critical constraints on the mass loss
history of the central stars. \htwo\ lines can potentially provide a powerful
probe of circumstellar nebulae around massive stars, as has been  demonstrated
for the case of $\eta$-Carinae, a well known LBV (Smith
\cite{2006ApJ...644.1151S}). However, \htwo\ lines around evolved massive stars
could be rare as \htwo\ might not survive in the vicinity of hot stars, and
might not be sufficiently excited around cool supergiants. Nevertheless, a
search for \htwo\ emission around evolved massive stars using UWISH2 survey data
might lead to new detections and could improve our understanding of their mass
loss history.

In our preliminary study, we searched for associated \htwo\ emission around
known LBVs and candidate LBVs from Clark et al. \cite{2005A&A...435..239C}.
Eight sources are covered by the UWISH2 survey, but all resulted in negative
detections. We also searched for \htwo\ counterparts of compact nebulae reported
by Gvaramadze et al. \cite{2010MNRAS.405.1047G}, most of which are proposed to
be associated with evolved massive stars. About 30 nebulae are covered by
UWISH2, and we found only one candidate with emission so far. Even though our
preliminary search has not been very successful, continuing search efforts in a
larger sample of evolved massive stars will be performed.

\subsection{Distances to star forming regions within our survey}

One major task of the survey will be to assign distances to the investigated
star forming regions. This is vital to achieve many of the science goals
discussed in the above sections. There are a number of possibilities:

The VLA H{\small I} GP survey, BU FCRAO $^{13}$CO Galactic Ring survey, and
Boston University-Arecibo Observatory H{\small I} survey can be used to assign
kinematic distances to each star forming region. H{\small I} absorption in the
aforementioned surveys can be used to resolve near/far distance ambiguities. 
Furthermore, the BGPS and Hi-GAL teams are working on using HISA, $^{13}$CO, and
other heterodyne observations of dense gas tracers (NH$_3$, HCO$^+$, N$_2$H$^+$,
CS, etc.) to determine distances to all of the dense gas clumps within the
northern galactic plane. Also, maser parallaxes to a number of massive star
forming regions are currently being measured (e.g. Reid et al.
\cite{2009ApJ...700..137R}).

Distances can also be estimated from mean extinction values in JHK colour-colour
diagrams, as well as counting the number of foreground stars in dense clouds and
comparing with the Besan\c{c}on Galaxy model (Robin et al.
\cite{2003A&A...409..523R}). It has been shown e.g. in Scholz et al.
\cite{2010MNRAS.406..505S} that this enables us to determine distances to not
too distant dark clouds with an accuracy of about 20\,\%. Finally, we can
utilise objects of known distance that are associated with the molecular clouds
we are investigating.

\section{Concluding Remarks}

The UWISH2 narrow-band imaging survey is being used to trace dynamic processes
associated with star formation and late stellar evolution. In particular, it
picks out active regions of star formation, leading to estimates of star
formation efficiency along the Galactic Plane. At the same time it also yields
a more complete, unbiased census of post-AGB stars and PN in the Milky Way,
leading to a comprehensive catalogue of PN morphologies as well as a map of
their distribution along the Inner Galaxy. 

A major strength of the UWISH2 survey lies in its complementarity with other
existing or upcoming surveys; the benefits of combining high-spatial resolution
narrow-band WFCAM images of massive star forming regions with mid- and far-IR
observations has recently been demonstrated by Davis et al.
\cite{2007MNRAS.374...29D} and Kumar et al. \cite{2007MNRAS.374...54K}. In
these complex environments the narrow-band data yield flow statistics and
pin-point regions of active star formation, while the multi-wavelength
photometry reveal the relative distributions of pre-stellar cores, protostars
and pre-main-sequence objects. 

Benefits are also wrought when \htwo\ data are combined with optical and radio
data, in outflows from low and high mass YSOs, but also toward post-AGB stars
and PN, where rapidly changing physical conditions require combined observations
in a variety of excitation tracers. The GLIMPSE-North data are already available
and accessible through the IPAC archive (and has been cross-matched to the
GPS, available at the WFCAM science archive); IPHAS data are also publicly
available and cover half of the Northern Galactic Plane; the AKARI all sky
survey faint point source catalogue will also be available in the near future.
The MSX point-source catalogue may also be used for the few massive sources that
saturate in IRAC data, along with future longer wavelength data from JCMT-JPS.
The importance of serendipitous discoveries of interesting emission-line objects
of all types should also not be ignored. 

The combination of broad- and narrow-band data will aid in the selection of
targets for study with instruments where full-scale mapping of the Galactic
plane is impractical (e.g. the sub-mm heterodyne array receiver, HARP, the
SCUBA-2 Polarimeter on the JCMT, and of course future high-resolution
facilities such as ALMA and the James Webb Space telescope). Finally, it is
worth noting that massive star formation is at the heart of the science case
for many Galactic Plane surveys, e.g. the JCMT Galactic Plane Survey (JPS),
the Spitzer-GLIMPSE survey, and the Herschel Hi-GAL mid-IR survey. These will
reveal dense molecular cores and mid-IR bright YSOs, effectively identifying
all of the massive pre-main-sequence stars, protostars and massive pre-stellar
cores. UWISH2 will in turn establish how many of these are dynamically active.
Although extinction will be high in the coldest and most massive objects,
outflows rapidly break out of these environments and are usually bright in
\htwo\ line emission. Identifying outflows and tracing them back to sub-mm
sources will certainly be an important way of identifying the location of
protostars in these mid-IR and sub-mm survey data. Indeed, our initial analysis
of 24 square degrees of early UWISH2 data, which has revealed numerous outflows
(as well as new PN, clusters and variable stars), demonstrates that this is
certainly the case.

\section*{acknowledgements}
 
We would like to thank the referee A.\,Ginsburg for his helpfull comments on the
paper. The UWISH2 survey team would also like to acknowledge the UKIRT support
staff, particularly the Telescope System Specialists (Thor Wold, Tim Carroll and
Jack Ehle) and the many UKIRT observers who have obtained data for the UWISH2
project via flexible scheduling. We also acknowledge the Cambridge Astronomical
Survey Unit and the WFCAM Science Archive for the reduction and ingest of the
survey data. BCK was supported by the Korean Research Foundation Grant funded by
the Korean Government  (KRF-2008-313-C00372). MSNK is supported by a Ci\^encia
2007 contract, funded by FCT/MCTES (Portugal) and POPH/FSE (EC). The United
Kingdom Infrared Telescope is operated by the Joint Astronomy Centre on behalf
of the Science and Technology Facilities Council of the U.K. Finally, we also
thank the UKIRT Time Allocation Committee for their support of this long-term
project.

\end{document}